\documentstyle[15lomcon,cite,epsfig]{article}

\begin{document}

\title{SEARCH FOR $\nu_{\tau}$ INTERACTIONS WITH THE NUCLEAR EMULSION FILMS OF THE OPERA EXPERIMENT}

\author{Fabio Pupilli \footnote{e-mail: fabio.pupilli@aquila.infn.it}, on behalf of the OPERA Collaboration}

\address{INFN - Laboratori Nazionali del Gran Sasso, I-67100 Assergi (L'Aquila), Italy}


\maketitle\abstracts{ The OPERA experiment aims at measuring the $\nu_{\mu} \rightarrow \nu_{\tau}$ oscillation through the $\nu_{\tau}$
appearance in an almost pure $\nu_{\mu}$ beam (CNGS). For the direct identification of the short-lived $\tau$ lepton,
produced in $\nu_{\tau}$ CC interactions, a micrometric detection resolution is needed.
Therefore the OPERA detector makes use of nuclear emulsion films, the highest spatial resolution tracking device,
combined with lead plates in an emulsion cloud chamber (ECC) structure called 'brick'. 
In this paper the nuclear emulsion analysis chain is reported;
the strategy and the algorithms set up will
be described together with their performances.}

\section{Introduction}

In the past decades several experiments provided compelling evidences supporting the neutrino oscillation hypothesis. For what concerns
the atmospheric neutrino sector, the disappearance of muonic neutrinos have been convincingly
observed in different experiments \cite{nuatmo}, exploiting both natural and artificial neutrino sources.
Nevertheless, the confirmation of the
$\nu_{\mu} \rightarrow \nu_{\tau}$ oscillation as the leading channel, through the direct observation of the $\nu_{\tau}$ appearance
is still missing.
The OPERA experiment \cite{proposal}, located in the underground Gran Sasso National Laboratory,
is aiming at the first direct observation of
neutrino oscil\-la\-tions in appearance mode by the identification of the $\tau$ lepton produced in $\nu_{\tau}$ CC interactions, in an
almost pure $\nu_{\mu}$ beam (CNGS \cite{cngs}) produced by the CERN SPS, 730 km far away from the
detector. The $\tau$ detection is accomplished by an event-by-event topological and kinematical reconstruction, and the micrometric
resolution needed to identify the short-lived $\tau$ particle is provided by nuclear emulsions.

\section{The detector}

OPERA is a hybrid detector \cite{OPERA} made of two indentical sections called Super Modules (SM), each one composed of a muon spectrometer and
a target section hosting the core of the apparatus, highly modular sandwiches of emulsion and lead plates (bricks).
The bricks are arranged in vertical structures called walls transverse to the beam direction; each wall is followed by a pair of tracker planes (TT)
made of plastic scintillator strips and providing bi-dimensional information used to identify the brick where the neutrino interaction
occurred and to generate the trigger. The muon spectrometer at the downstream end of each SM allows measuring charge and momentum
of penetrating tracks.

\section{OPERA bricks and emulsion automatic scanning}

\begin{figure}[t]
\centering
\includegraphics[height=2.2cm]{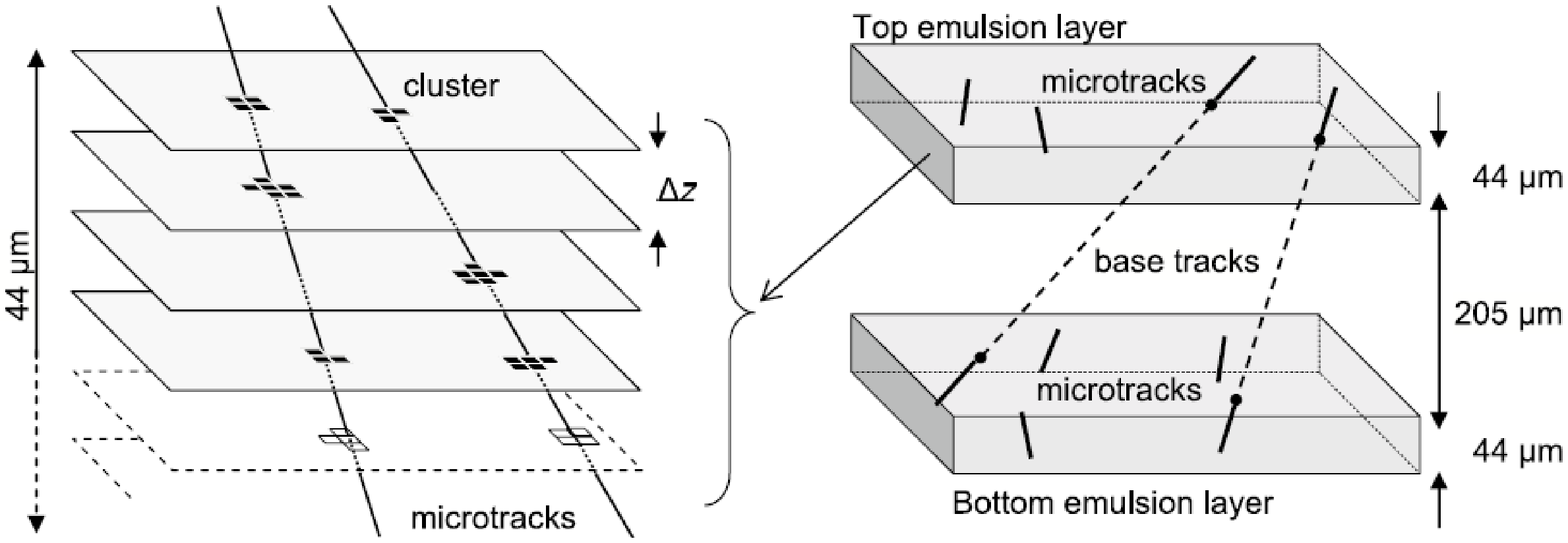}
\caption{Micro-track connection across the plastic base (base-track)}
\label{fig:mtr-btr}
\end{figure}

The OPERA bricks exploit the ECC (Emulsion Cloud Chamber) technique: emulsion films are interspaced with lead plates, providing
high tracking resolution and large target mass in a modular way.
Each brick is composed of 57 emulsion films interleaved with 56 lead plates of 1 mm thickness.
The transverse dimension of a brick are 12.5 $\times$ 10.0 cm$^{2}$ and the thickness along the beam direction is 7.9 cm, corresponding
to about 10 radiation lengths. The weight is about 8.3 kg. A tightly packed removeable doublet of emulsion films, called Changeable Sheets
(CS) \cite{CS}, is attached to the downstream face of each brick; CS act as offline ``triggers" and interfaces between the TT and the brick, with the
task of validating the result of the algorithm for the selection of bricks and allowing the transition from the centimeter resolution of the
electronic detectors to the micrometric resolution of nuclear emulsions. A total amount of 150000 bricks have been produced and placed
in the walls of the detector, for a total target mass of $\sim$ 1.25 ktons; this translates in an overall emulsion surface larger
than 100000 m$^{2}$. The industrial production of such a large amount of films was carried out, after a joint R\&D program \cite{emulsion},
by the Fuji Film company. An OPERA film has 2 emulsion layers (each 44 $\mu$m thick) on both sides of a transparent triacetylcellulose
base (205 $\mu$m thick). The total thickness is 293$\pm$5 $\mu$m; the sensitivity of a film amounts to $\sim$ 36 grains/100 $\mu$m
for minimum ionising particles.
Given the rate of $\sim$ 20 neutrino in\-te\-rac\-tions/day and the corresponding large emulsion surface to be analysed, a dedicated R\&D
project was pursued in OPERA leading to the development of fully automated high-speed scanning systems \cite{microscope}.
The systems are based on a motorised stage, holding the emulsion film, and on an optics equipped with a camera,
moving along the Z direction and focusing different depths in emulsion: in this way, while the area to be scanned is spanned in the X-Y
direction, the camera produces optical tomographic image sequences of each emulsion layer.
The acquired images are digitized and processed by a control workstation to reconstruct aligned sequences of grains through an
emulsion layer ({\it micro-tracks}). The linking of two matching micro-tracks  produces the so-called {\it base-track} (see Figure \ref{fig:mtr-btr});
this reduces the instrumental background due to fake combinatorial alignments. The analysis of an emulsion stack continues with
the film-to-film alignment, computing the six parameters of an affine transformation by a suitable alignment pattern (as described
in the following section), and with the {\it volume-track} formation by fitting aligned base-tracks in the analysed volume.
The performances of this analysis procedure have been evaluated in a 10 GeV pion exposure
of an OPERA emulsion stack without lead, to test intrinsic resolutions and efficiencies
\cite{scanning-res}: the position resolution is sub-micrometric, while the angular resolution is of the order of a mrad;
the base-track finding efficiency is on average $\sim$ 90\%.

\section{Event analysis}

Once signals in the electronic detectors are compatible with an interaction
inside a brick, data are processed by a software reconstruction program
that selects the brick with the highest probability to contain the neutrino interaction vertex \cite{eledet}.
The impact point on CS of the muon (for CC-like events) or of the barycenter of the shower (for NC-like events) is also predicted.
The brick is extracted from the detector and is stored in a shielded area underground, waiting for confirmation by the CS films
analysis. Given the measured position resolution of the TT reconstruction ($\sim$ 8 mm), a large area around the
prediction is scanned on each CS film; the films are roughly aligned exploiting four X-ray marks printed before the
development of the doublet (with a precision of $\sim$ 10 $\mu$m) and finely aligned using tracks from low energy electrons emitted in
the decay of natural radioactive isotopes ($\sim$ 2 $\mu$m precision). If tracks compatible with an interaction in the
brick and with electronic detector data are found in the CS films, the corresponding brick undergoes an exposure to X-rays, to produce lateral
marks on the film edges for a coarse alignment, and to high energy cosmic rays, to provide a pattern of tracks for a refined
film-to-film alignment; then it is dismantled and the emulsion films are de\-ve\-lo\-ped. If no signal is detected in the CS films,
the brick is reinserted in the detector with a fresh CS doublet and the next most probable brick is extracted.
All tracks found in the CS films are looked for in the most downstream film of the brick and then followed back from film to film ({\it scan-back}),
exploiting the lateral X-ray mark alignment, until they are not found in five consecutive films. The stopping point is considered
as the signature either for a primary or a secondary vertex. The vertex confirmation is performed by scanning an area
of 1 cm$^{2}$ on 5 upstream and 10 downstream films with respect to the stopping point ({\it volume-scan}).
All base-tracks are reconstructed and aligned
with a micrometric precision, thanks to the cosmic ray alignment tracks, and the volume tracks are constructed. The interaction point is identified by
an automated algorithm as the minimum distance point of two or more tracks and the vertex is fully reconstructed. A dedicated procedure,
called {\it decay search}, is performed with the aim of finding eventual decay topologies, secondary interactions or $\gamma$-ray
conversions, sear\-ching for large kink angles or for tracks with large impact parameter (IP)
with respect to the vertex (since the IP for primary tracks is not larger than 10 $\mu$m). When a secondary vertex is found, a kinematical
analysis is performed, exploiting the high resolution ensured by nuclear emulsions: momenta of charged particles are estimated through
the angular deviations due to the Multiple Coulomb Scattering of particles in lead plates (with a resolution of $\sim 22\%$) \cite{momentum};
the energy of electromagnetic showers is measured by a Neural Network shower shape analysis, which takes into account also the
Multiple Coulomb Scattering of the leading tracks. 
This analysis chain led to the reconstruction of $\sim$ 2800 neutrino interactions in the first two physics runs \cite{2008-2009},
and to the observation of a first $\nu_{\tau}$ candidate event \cite{event}.

\section{Conclusions}

Thanks to their high resolution, nuclear emulsions play a key role for the purposes of the OPERA experiment.
High speed automatic scanning systems and dedicated procedures have been developed
for each analysis step, to ensure a micrometric resolution in position and an angular resolution of the order of a mrad, needed
for the $\tau$ decay topological identification. Nuclear emulsion films are also used in kinematical analysis of decay topologies.
The OPERA nuclear emulsion analysis chain has proven to be successful, leading to the reconstruction of $\sim$ 2800
neutrino interactions and of a first $\nu_{\tau}$ candidate.


\section*{References}
\begin{thebibliography}{99}

\bibitem{nuatmo} Y. Fukuda {\it et al.} [Super-K Coll.], {\it Phys. Rev. Lett.} {\bf 81}, 1562 (1998). \\
                 M. H. Ahn {\it et al.} [K2K Coll.], {\it Phys. Rev.} {\bf D 74}, 072003 (2006). \\
                 P. Adamson {\it et al.} [MINOS Coll.], {\it Phys. Rev. Lett.} {\bf 101}, 131802 (2008).

\bibitem{proposal} M. Guler {\it et al.} [OPERA Coll.], {\it CERN-SPSC-2000-028}.

\bibitem{cngs} CNGS project: http://proj-cngs.web.cern.ch/proj-cngs/.

\bibitem{OPERA} R. Acquafredda {\it et al.} [OPERA Coll.], {\it JINST} {\bf 4}, P04018 (2009).

\bibitem{CS} A. Anokhina {\it et al.} [OPERA Coll.], {\it JINST} {\bf 3}, P07005 (2008).

\bibitem{emulsion} T. Nakamura {\it et al.}, {\it Nucl. Instrum. Meth.} {\bf A 556}, 80 (2006).

\bibitem{microscope} L. Arrabito {\it et al.}, {\it Nucl. Instrum. Meth.} {\bf A 568}, 578 (2006). \\
                     K. Morishima and T. Nakano, {\it JINST} {\bf 5} P04011 (2010).

\bibitem{scanning-res} L. Arrabito {\it et al.}, {\it JINST} {\bf 2}, P05004 (2007).

\bibitem{eledet} N. Agafonova {\it et al.} [OPERA Coll.], {\it New J. Phys.} {\bf 13}, 053051 (2011). \\
                 R. Rescigno, {\it The electronic detectors of the hybrid OPERA neutrino experiment}, these proceedings.

\bibitem{momentum} N. Agafonova {\it et al.} [OPERA Coll.], arXiv:1106.6211v1 [physics.ins-det], submitted to {\it New J. Phys}.

\bibitem{2008-2009} N. Agafonova {\it et al.} [OPERA Coll.], arXiv:1107.2594v1 [hep-ex].

\bibitem{event} N. Agafonova {\it et al.} [OPERA Coll.], {\it Phys. Lett.} {\bf B 691}, 138 (2010).


\end{thebibliography}
\end{document}